\documentclass{emulateapj}

\shorttitle{X-Ray Periodicity from M82} 
\shortauthors{Kaaret, Feng, \& Lang}

\begin{document}

\title{Confirmation of the 62 Day X-Ray Periodicity from M82}

\author{Philip Kaaret and Hua Feng} 
\affil{Department of Physics and Astronomy, University of Iowa,  Van
Allen Hall, Iowa City, IA 52242.}


\begin{abstract}

Using 400 days of new X-ray monitoring of M82, we confirm the 62~day
periodicity previously reported.  In the full data set spanning
1124~days, we find a period of $62.0 \pm 0.3$ days and a coherence, $Q =
22.3$, that is consistent with a strictly periodic signal.  We estimate
that the probability of chance occurrence of our observed signal is $6
\times 10^{-7}$.  The light curve folded at this period is roughly
sinusoidal and has a peak to peak amplitude of $(0.99 \pm 0.10) \times
10^{-11} \rm \, erg \, cm^{-2}\, s^{-1}$.  Confirmation of the
periodicity strengthens our previous suggestion that the 62~day
modulation is due to orbital motion within an X-ray binary.

\end{abstract}


\section{Introduction}

Bright, non-nuclear X-ray sources in external galaxies, the so-called
ultraluminous X-ray sources (ULXs), represent either intermediate-mass
black holes \citep{Colbert99,Makishima00,Kaaret01} or super-Eddington
accretion onto stellar-mass black holes.  The brightest X-ray source in
the nearby starburst galaxy M82,  CXOU J095550.2+694047 = X41.4+60
\citep{Kaaret01}, is one of the most extreme ULXs.  Assuming isotropic
radiation, a black hole mass of at least $500 M_{\sun}$ is required to
avoid violating the Eddington limit.  The source also shows
quasiperiodic oscillations (QPOs) at relatively low frequencies,
50-190~mHz, suggesting a relatively high compact object mass
\citep{Strohmayer03,Dewangan06,Mucciarelli06,Kaaret06b}.  The relative
proximity and brightness of source enables studies that are not feasible
for other ULXs.

We monitored the X-ray emission from M82 for 240~days in 2004/2005 and
detected a period of 62~days \citep{Kaaret06a}.  We interpreted the
62~day period as the orbital period of the ULX binary system.  For a
system in Roche-lobe contact, such a long orbital period implies a low
density companion on the giant or supergiant branch.  Identification of
the evolutionary phase of the companion star represents a significance
advance in our knowledge of ULXs.  In order to test this interpretation
of the 62~day periodicity, we obtained additional monitoring of M82 in
2006/2007.  We describe the new X-ray observations in \S~2 and discuss
the results in \S~3.

\section{Observations}

We obtained 187 observations of M82 using the Proportional Counter Array
(PCA) on the Rossi X-Ray Timing Explorer (RXTE) covering MJD 53825 to
54223 with observations approximately every other day under RXTE program
92098.  The observations were typically less than 1~ks.  We also
analyzed archival data from RXTE program 90121 which consists of 144
observations roughly every other day from MJD 53252 to 53490 and two
observations made earlier. In this earlier program, the observations
were typically about 2~ks each.

\begin{figure*}[tb]
\centerline{\includegraphics[width=6.25in]{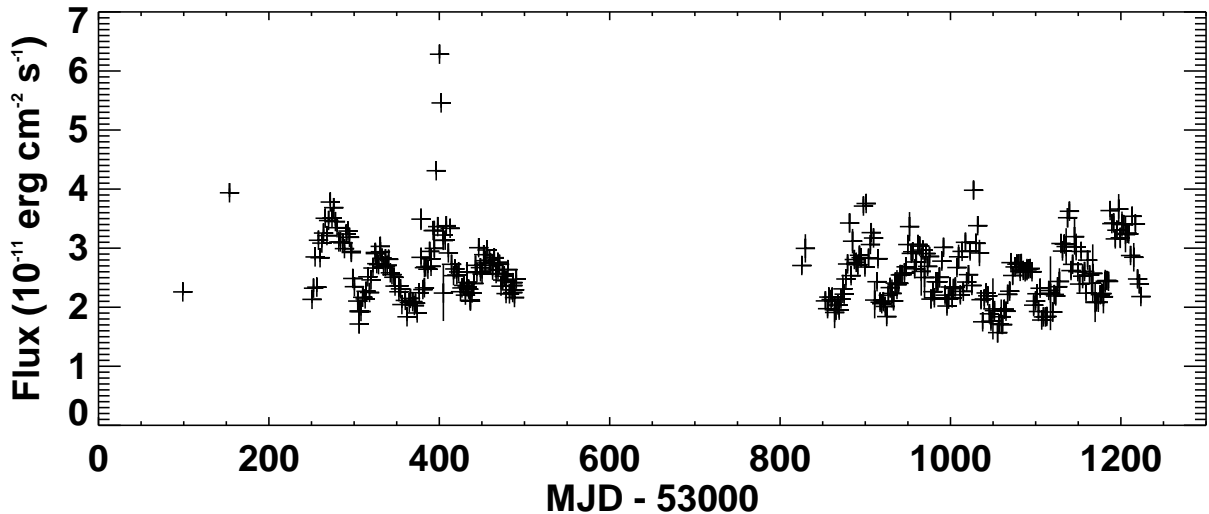}}
\caption{\label{lightcurve} Light curve of M82 in the 2--10~keV band.
The plot shows the flux measured using the PCA for each observation
versus the observation date in MJD.  The flux includes contributions
from all X-ray sources within M82.} \end{figure*}

We used the RXTE production data and processed the data using HEAsoft
version 6.2.  We selected good time intervals where the time since the
last SAA passage was more than 30 minutes, the electron contamination
was less than 0.1, and the pointing was within $0.1\arcdeg$ of the
target and at least $10\arcdeg$ above the horizon.  We produced spectra
using only the top layer in Proportional Counter Unit 2 and estimated
the background using the faint source background model.  We fitted the
spectra using XSPEC 11 in the energy range 2.6--20~keV with a power-law
model with an interstellar absorption column density fixed to $3 \times
10^{22} \rm \, cm^{-2}$.  The best fit model was used to calculate the
absorbed flux in the 2-10 keV band.

Fig.~\ref{lightcurve} shows the flux in the 2-10~keV band versus time. 
We note that, due to the large angular acceptance of the PCA, the
measured flux includes contributions from all X-ray sources within M82.
Thus, some part of the fluctuations represents sources other than
X41.4+60.  There is an X-ray flare around MJD 53400 which was discussed
in \citet{Kaaret06b}.  Points with fluxes above  $4 \times 10^{-11} \rm
\, erg \, cm^{-2} \, s^{-1}$ were removed in the subsequent analysis. 
The photon index is generally between 2.0 and 2.7 with an average value
of 2.4.  The distribution of the measured photon indexes are consistent,
within the uncertainties, with a constant value of 2.4.

\begin{figure}[tb]
\centerline{\includegraphics[width=3.25in]{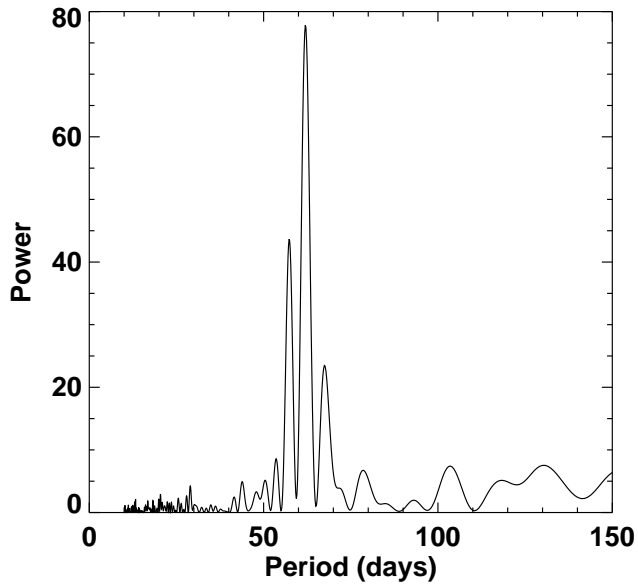}}
\caption{\label{periodogram} Periodogram of the 2--10~keV light curve of
M82.  The strongest peak is at a period of $62.0 \pm 0.3$~days.  The two
secondary peaks near the main peak are aliases due to the gap in the
monitoring.  The powers are calculated using the method of
\citet{Horne86} and are normalized by the total variance of the fluxes.}
\end{figure}

The light curve shows an apparent modulation with a period near
60~days.  Fig.~\ref{periodogram} shows a periodogram with the power
normalized by the total variance of the data \citep{Horne86}.  There is
a peak at a period 62.0~days with a power of 77.9.  We estimate the 90\%
confidence error on the period to be 0.3~days.  There are two secondary
peaks near the main peak which are aliases due to the gap in the
monitoring.  Retaining the X-ray flare near MJD 53400 does not shift the
peak, but decreases the power (because of the normalization to the total
variance of the data) to 70.5.  Using fluxes calculated from spectral
fits with the photon index fixed to 2.4 does not significantly change
the period or the power of the peak.

We tested the significance of the observed signal using a red noise
background, as is appropriate for an accreting X-ray source
\citep{Israel96,Vaughan05}.  We fitted the power versus frequency
relation for periods in the range 6--280 days to a power-law form and
found a spectral index of $-1.04 \pm  0.06$.  We generated red noise
with a spectral index of $-1.10$ and with mean and variance equal to
those of the data using the {\tt rndpwrlc} routine of the {\tt aitlib}
IDL subroutine library provided by the Institut f\"ur Astronomie und
Astrophysik of the Universit\"at T\"ubingen \citep{Timmer94}.  The
duration of each generated light curve is longer than the actual data in
order to minimize the effects of red noise leakage.  Each light curve
contains 8192 data points with uniform spacing of 0.66~days and a subset
of 330 points from the middle of this set with relative times matching
the actual observations are extracted for analysis.  These 330 simulated
fluxes were processed with the same procedures used to analyze the real
data.  We generated $2 \times 10^{6}$ trial light curves and searched
for cases where the power at periods of 10 to 150 days was greater than
or equal to the observed value of 77.9.  We found one such case and
estimate the probability of chance occurrence of our observed signal to
be $5 \times 10^{-7}$.  Fitting the distribution of maximum observed
powers, we estimate that the probability of chance occurrence of our
observed signal is $6 \times 10^{-7}$, in good agreement.  This
procedure is conservative because it includes the signal in the
calculation of the power-law slope and the variance and because the
period search range, 10--150 days, extends to significantly lower
frequencies than the observed period where the red noise produces high
amplitude fluctuations.  If we restrict the search range to periods of
62 days or less, then a fit to the distribution of the maximum observed
powers indicates that the probability of chance occurrence of our
observed signal is $3 \times 10^{-13}$.

The coherence or quality value of the peak signal, the period of the
peak divided by the full width at half maximum power, is $Q = 22.3$. 
This is fully consistent with that expected for a periodic process given
the observation duration.

\begin{figure}[tb]
\centerline{\includegraphics[height=3.25in]{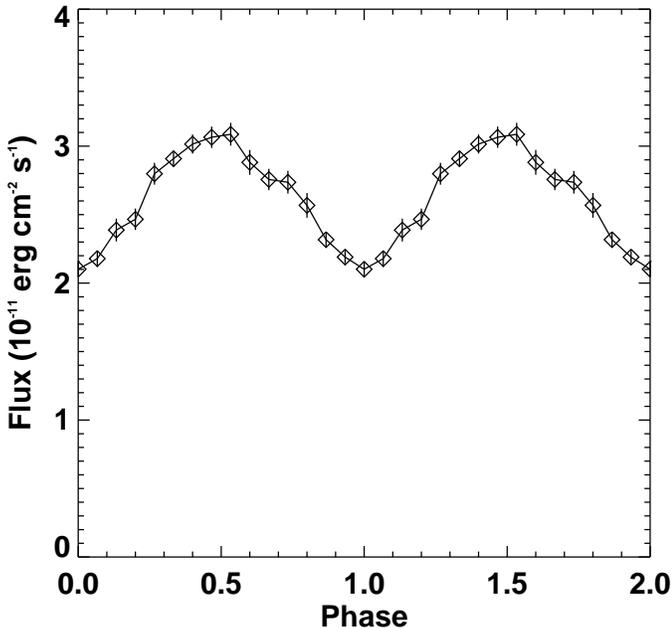}}
\caption{\label{phase} Flux from M82 folded at a period of 61.98 days. 
Two periods are shown for clarity.  Each point is the average flux of
the observations falling within the given phase bin and the error bar is
the standard error.} 
\end{figure}

Fig.~\ref{phase} shows the data folded at the best fit period.  Each
point is the average flux of the observations falling within the given
phase bin and the error bar is the standard error, i.e.\  the standard
deviation of the fluxes in each bin divided by the square root of the
number of fluxes in that bin.  The amplitude of the modulation, taken as
the maximum average flux in one bin minus the minimum, is $(0.99 \pm
0.10) \times 10^{-11} \rm \, erg \, cm^{-2} \, s^{-1}$.

To search for rapid variability, we extracted event files with high time
resolution data for the 187 observations.  Events in the 2.4-11.9~keV
energy band were selected in the good time intervals defined above and
split into segments of 256~s.  An FFT with a time resolution of 1~s was
calculated for each segment.  The FFTs within each observation were
added incoherently.  The resulting total power spectrum was
logarithmically rebinned with a bin width of 16\%, equal to the widths
of the QPOs previously detected from M82.  We searched for individual
bins with high powers and calculated the significance taking into
account the number of bins in each power spectrum.  The highest power
was recorded on MJD 54155.148 in a single 256~s interval of data at a
frequency of $20 \pm 4$~mHz with a Leahy power of 19.7 corresponding to
a chance probability of occurrence of $5.4 \times 10^{-5}$ ($4.0\sigma$)
single trial and 0.0012 taking into account the trials for that single
observation.  Taking into account all observations in program 92098, the
chance probability of occurrence is 0.23 indicating that the QPO
detection is not significant.  The observations in the new RXTE program
are too short to provide good sensitivity for QPO detection.

\section{Discussion}

Using 400 days of new X-ray monitoring of M82, we confirm the 62~day
previously reported \citep{Kaaret06a}.  The coherence of the signal, $Q
= 22.3$, is consistent with a strictly periodic signal.   The light
curve folded at this period is roughly sinusoidal and has a peak to peak
amplitude of $1.0 \times 10^{-11} \rm \, erg \, cm^{-2}\, s^{-1}$.  The
new observations rule out the possibility that the signal could be
red-noise fluctuation.  The remaining possible interpretations of the
signals are that it represents either the orbital period of the binary
system or a superorbital modulation.

Superorbital modulations are most often interpreted as due to accretion
disk precession \citep{Wijers99,Ogilvie01} and, in this case, would
represent variations in our viewing angle of the disk.  Indeed, in the
one source where both a precessing relativistic jet (thought to be
launched perpendicular to the disk) and a superorbital modulation are
detected, SS 433, the periods of jet precession and superorbital
modulation are the same \citep{Margon89}.  If ULXs are beamed sources,
then the beaming cone would most naturally be perpendicular to the disk
axis.  Thus, accretion disk precession would naturally produce
modulation of the beam at the superorbital period.

\begin{figure}[tb] \centerline{\includegraphics[height=3.25in]{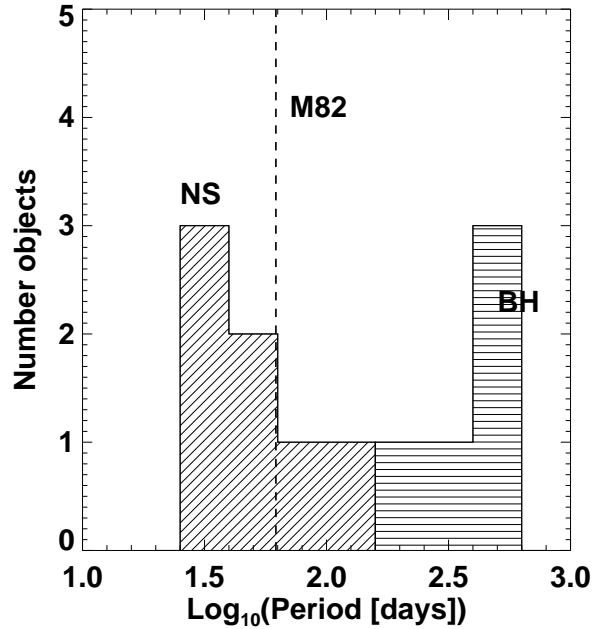}}
\caption{\label{superorb} Distribution of superorbital periods of
neutron star binaries (diagonally hatched) and black hole candidate
binaries (horizontally hatched).  The vertical dashed line shows the
period measured from M82; it falls in the neutron star binary range.}
\end{figure}

The distribution of superorbital periods of neutron star and black hole
candidate X-ray binaries are plotted in Fig.~\ref{superorb} 
\citep{Wijers99,Smith02,Corbet03,Rau03}.  We include only black hole
candidates as defined by \citet{Remillard06}.  It is apparent from the
figure that there are no black hole binaries with superorbital periods
near 62 days.  The shortest superorbital period from a dynamically
confirmed black hole candidate is the 198~day period of LMC X-3 which is
longer by more than a factor of 3.  Thus, if the 62 period from M82 is
interpreted as a superorbital period, this would suggest that it arises
from a neutron star system.  The measured flux modulation implies a
luminosity (assuming isotropic emission) of $1.6 \times 10^{40} \rm \,
erg \, s^{-1}$ for a source in M82 at a distance of 3.63~Mpc.  This
would exceed the Eddington limit for a $1.4 M_{\odot}$ neutron star by
at least a factor of 86.  The brightest known neutron star transient is
A0538-66 which reached a peak luminosity of $8\times 10^{38} \rm \, erg
\, s^{-1}$ \citep{White78}, a factor of 20 lower than X41.4+60. 
Compared instead to the peak luminosity of flares from X41.4+60, the
peak luminosity of A0538-66 is a factor of 95 lower.  Thus, a neutron
star interpretation for X41.4+60, as would be expected if the 62 day
period is a superorbital modulation, is untenable.

When monitored over many periods, superorbital modulations show reduced
coherence in the form of period or phase shifts.  The high coherence
measured for the M82 periodicity is inconsistent with all but the most
stable of the superorbital modulations, specifically that of Her X-1.

Several X-ray binaries produce X-rays modulated at the orbital period
including Cygnus X-3 \citep{Elsner80}, LMC X-3 \citep{Boyd01}, 1E
1740.7-2942 and GRS 1758-258 \citep{Smith02}, and GX 13+1
\citep{Corbet03}.  Thus, the periodicity from M82 may be interpreted as
due to orbital modulation.  The  coherence of the signal we observe from
M82 is consistent with that expected for a strictly periodic signal
given the observational coverage.  This is consistent with
interpretation as an orbital modulation.   We conclude that the 62~day
periodicity most likely indicates the orbital period of an X-ray binary.
If the companion fills it Roche-lobe, as expected in a system with a
mass accretion rate high enough to produce the observed X-ray flux even
with moderate beaming, then the long period indicates that the companion
star has a low average density, $5 \times 10^{-5} \rm \, g \, cm^{-3}$,
and is therefore a giant or supergiant.

\acknowledgements

PK and HF acknowledge partial support from NASA Grant NNX06AG77G.  PK
acknowledges support from a University of Iowa Faculty Scholar Award.


\label{lastpage}

\end{document}